\documentclass[letterpaper,twocolumn,10pt]{article}
\usepackage{usenix2024_SOUPS}
\usepackage{tikz}
\usepackage{amsmath}

\begin{document}

\date{}

\title{\Large \bf Traces of Abuse: How Generative AI Impacts Image-Based Sexual Abuse (IBSA) Investigations}

\def\plainauthor{Wyss et al.}

\author{
{\rm Jasmin Wyss}\\
Ruhr University Bochum
\and
{\rm Ivy Turk}\\
Ruhr University Bochum
\and
{\rm Anna Neumann}\\
University Duisburg-Essen\\Research Center for Trustworthy AI
\and
{\rm  Rebekah Overdorf}\\
Ruhr University Bochum\\Research Center for Trustworthy AI
} %

\maketitle
\thecopyright

\begin{abstract}
The introduction of generative AI (GAI) into the workflow of image-based sexual abuse (IBSA) only worsened the ease of creation and distribution, victimizing more people than ever.  
We outline how the introduction of generative AI (GAI-IBSA) impacts the creation of traces and the type of reasoning they allow. We illustrate the impact by comparing the forensic traces available in four different IBSA scenarios. We discuss the impacts on the (possibility of) investigation, arguing that the advent of generative AI overall benefits abusers by making perpetration easier, and the perpetrator harder to trace.
\end{abstract}

\section{Introduction}\label{sec:into}

Image-Based Sexual Abuse (IBSA) is a cyber-enabled offense arising from offenders having access to sensitive media. In the digital era, disseminating IBSA has become effortless: social media enables the spreading of IBSA content, and the structure of the internet facilitates anonymity for the perpetrators. These affordances also exacerbate the harms caused by IBSA by making private content irrevocably public and widely available. The low barrier of entry to perform IBSA has contributed one in five people being victimized by IBSA~\cite{umbach25ibsavictimisation}. 

Still, until recently, the barrier to carrying out IBSA has been access to intimate media: Intimate imagery may be consensually shared through communication with an intimate partner and then later non-consensually distributed, or it may be obtained through physical access to the survivor or their devices. A perpetrator without direct access to intimate imagery can also create it using photo editing tools, yet this task is time-consuming and requires extensive photo editing skills.

AI-generated non-consensual sexual imagery, a subset of what are commonly known as ``deepfakes'', lowers this barrier considerably. Modern media generation models can, for example, ``nudify'' or ``bikinify'' an image by artificially removing clothing from a photo and simulating the body underneath~\cite{gibson2025analyzing, wiredGrokStill, 404mediaElonMusks}. This process means that IBSA mediated by generative AI (GAI-IBSA) no longer requires an offender to learn photo editing skills and create an image using traditional photo editing tools or to otherwise obtain intimate imagery. Instead, they can take any photo of a person, including non-intimate photos from social media, and AI-generate deepfaked intimate media.

Any new technology that impacts the implementation of a offense necessarily impacts the traces that are left behind by the execution of the offense. In forensic science, \textit{traces} are remnants of a presence or activity~\cite{Sydneydeclaration, ribaux2023police}. They aid in understanding how an offense was committed and who perpetrated the offense.

The shift from IBSA to AIG-IBSA impacts the digital \textit{traces} that are left behind by a perpetrator that aid in identification. In this position paper, based on four general scenarios of IBSA realizations, we discuss how generative AI impacts the available traces, and therefore the broader impact on victimization via IBSA.

\section{Background and Related Work}\label{sec:background}

We first survey the related literature on IBSA and explain generally the concepts of forensic analysis and traces. 

\subsection{IBSA}\label{sec:ibsa}
Image-Based Sexual Abuse (IBSA) is the non-consensual creation and/or (threatened) distribution of sexually explicit images or videos, often without the victims' consent or knowledge~\cite{mcglynnImageBasedSexualAbuse2017}. It manifests in different forms, including Non-Consensual Intimate Images/Media (NCII/NCIM, colloquially known as ``revenge porn''), sextortion, upskirting/downblousing, and cyberflashing~\cite{mcglynn25beyond}. It is most commonly perpetrated by (ex) partners, but perpetrators can also be (former) friends, family members, coworkers or strangers~\cite{ruvalcaba2020nonconsensual}. %

IBSA affects many people. A survey of 16\,000 people by Umbach et al.~\cite{umbach25ibsavictimisation} found that 22.6\% of respondents had experienced IBSA at some point in their lives. Notably, LGBTQ+ participants (38.5\%) and young people (33\%, under 35) were more likely to be victimized than other groups. The impact on victims is extensive as detailed by Champion et al.~\cite{champion22impacts}. There are mental health consequences such as anxiety, stress, and depression, which have knock-on impacts on victims' ability to perform daily activities. Work and academic life is impacted through inability to focus on tasks and the triggers that are present in seemingly mundane interactions. Victimization also leads to mistrust in strangers and can impact how survivors behave online. 

\subsubsection{Terminology}
Terminology for IBSA and related topics varies by research field. Intimate imagery may instead be referred to as intimate \textit{media}, highlighting that videos and other formats may be used in addition to photographs. The term ``intimate'' may be replaced with ``(sexually) explicit'', and a distinction may be drawn between the offenses of \textit{creating} and \textit{distributing} content. Together, this means that non-consensual sharing of intimate media may be referred to as NCII, NCIM, NCDI, and NCEI.

For artificial content, researchers may prefix these terms with ``Artificial Intelligence Generated'' (\textbf{AIG}-NCII) or insert the term ``synthetic'' (NC\textbf{S}EI). Colloquially, artificially generated content is called ``deepfaked,'' though on its own this term is broader than IBSA. 

When referring to the person who is subject to this offense, criminology traditionally defines them as ``victims'' based on the study of victimization. However, to empower those who have suffered this abuse, support organizations have adopted the terms ``survivor'' and ``victim-survivor''. This is opposed by some as it implies helplessness and the need to be saved~\cite{ahmed26critique}, with preference given to community-defined terms. However, for lack of a better term at the time of writing, we choose the term ``survivor'' to describe those that have experienced IBSA.

\subsubsection{Abuse Environments}
IBSA can arise in several scenarios. In Intimate Partner Abuse (IPA), domestic abusers often have extensive access to their partners' devices~\cite{freed18AStalkersParadise, doerfler2024privacyvstransparency}, which facilitates the gathering of intimate images. Furthermore, partners may consent to the recording of intimate activities, but not to their distribution by the abuser at a later date.

Online dating apps also facilitate IBSA. Online dating apps center images as the precursor to user interactions, often forcing image and personal data disclosure as a prerequisite to participation. They also provide opportunity, with many potential victims concentrated in one platform, and remove bystanders who may otherwise deter offenses.~\cite{wolbers2024routine}
 
\subsection{Forensic Analysis}\label{sec:forensic}

Traces are the central subject of forensic science. 
Forensic science is defined by Roux et al. as ``a case-based (or multi case-based) research-oriented, science-based endeavor to study traces --- the remnants of past activities (such as an individual’s presence and actions) --- through their detection, recognition, recovery, examination and interpretation to understand anomalous events of public interest (e.g., crimes, security incidents)''~\cite{Sydneydeclaration}.

During the \textit{investigative process}, traces are typically used to reduce the suspect pool, identify potential suspects that need to be investigated further, or to reconstruct events. 

If used for \textit{evaluative purposes}, the observed traces are used to support or reject hypotheses related to the case. Typically, hypotheses are about the source of a trace (this abusive image was created by that software), an activity that left a specific trace, or, more rarely, about a criminal action at the origin of the trace. As an example, trace evidence can be used to link a suspect's device to the creation of NCSEI if the traces on the device\footnote{These may include NCSEI videos on the device, installed software that performs nudification, or reference material, e.g. images of the victim} are more likely to occur if a user of the device created NCSEI material, rather than if the device was not used for the creation of the material. 

Independent from either reducing the suspect pool or evaluating the strength of the evidence, traces can also be used to \textit{create intelligence}. For example, if one perpetrator cyberflashes multiple victims, the image material can be used to link the cases to the same perpetrator.

\section{Scenarios}\label{sec:scenarios}
In this section, we analyze how the advent of generative AI impacts different types of IBSA. For each form of abuse, we provide a definition of the general offense, then discuss how the abuse is enacted \textit{before} the introduction of deepfakes based on an example. Then we describe how \textit{generative AI} allows perpetrators to perform the offense differently and the \textit{consequences} of this change. Our examples are not exhaustive; rather, they are intended to demonstrate the transformative power of generative AI in cases of IBSA. However, we base the examples on previous research and reported cases, including descriptions of how a case of IBSA might unfold.

\subsection{NCEI}\label{sec:revengeporn}
Non-consensual explicit imagery (NCEI) describes acts of non-consensually creating, retaining or sharing explicit imagery, often to enact revenge against or control over the depicted person~\cite{wei24ibsahelp}. One prevalent form of NCEI is a former partner publishing explicit imagery of the survivor to a wider audience.

\medskip
\noindent\textbf{Before} the advent of deepfakes, the creation of the imagery in question required the survivor to first be recorded while engaging in the depicted acts. Then, the perpetrator, if not the person creating the imagery, needed to get access to the material. Lastly, they needed a way of making the material available to others. 
\textit{In an example}, Mallory covertly films themselves and Alice engaging in sex while they are in a relationship. After Alice ends the relationship, Mallory uploads the recordings to a public video-sharing platform in retaliation.

When Alice learns of the video, they are able to deduce that the video was made by Mallory and can report them to the authorities. 
Mallory is identified as the culprit based on contextual information depicted on film (the video having been filmed in their shared apartment), embedded metadata of the video, and parts of Mallory's body (such as tattoos) being visible in some sections. Furthermore, during the investigation, the police can obtain data from the platform provider about the account that shared the video. This data includes a collection of IP addresses associated with log-in times and an email address used for creation of the account. 

As such, the only valuable information provided by data collection from the platform is indications of the time-zone the perpetrator might have been in given aggregated login times. 
If law enforcement searches Mallory's devices, they might find traces useful for evaluative purposes: a version of the video that predates the upload, or traces indicating that the associated uploading account was used on the device, e.g. the username and password are saved.

\medskip
\noindent\textbf{With generative AI}, the creation and access to explicit imagery of a person no longer necessitates said person to have been captured, i.e. photographed or filmed, engaging in the depicted acts. If the perpetrator has access to any reference images of the victim(s) and is able to access and prompt a media-generating model effectively, the perpetator can create synthetic explicit imagery.

\textit{In our example}, Mallory can use non-explicit images of their former partner Alice to create the non-consensual explicit imagery they later share on a public platform. The effects of the publication of the video on Alice are similar, i.e. public humiliation and inability to delete the shared material~\cite{umbach25ibsavictimisation, champion22impacts, Rigotti_McGlynn_Benning_2024}. 
However, apart from the timing of the upload, there is nothing about the video that indicates that Mallory is the creator. Even the timing is only circumstantial, as Alice has shared multiple images and videos of themselves online. As such, everybody with an internet connection has 
enough reference material to create the video. The only traces that remain are personally identifiable information about the account creator and time of upload.
If Mallory's phone is searched based on other evidence, then the relevant traces are indications of synthetic material creation, including deepfake/nudification software, reference material of Alice, a copy of the generated video, or connections to the video-uploading account.

\medskip
\noindent\textbf{Consequences:} As illustrated in the example, it is harder to identify who created the AI-generated non-consensual material. Similar dynamics apply if existing NCEI was ``only'' AI-\textit{modified} rather than fully AI-\textit{generated}. For example, the perpetrator is most likely not depicted, either due to generation of a non-existent person or modifying the image to be anonymous. Additionally, the metadata can only indicate if a video is generated through a metadata-level watermark~\cite{fernandez2025a}.
While there are potentially new traces associated with the creation of the video, such as additional accounts on an AI-video generation platform or service, they cannot be easily accessed or known to exist. In many cases, for investigators to access useful traces, they need to be able to search the suspects' devices.

\subsection{Sextortion}\label{sec:sextortion}
Sextortion is a form of extortion where the perpetrator uses the threat of publishing intimate images of the survivor to coerce them into complying with their demands.

\medskip
\noindent\textbf{Before} the advent of easily accessible generative AI methods, %
sextortion would typically comprise of two phases: First, the intimate image of the survivor is obtained through an exchange between survivor and perpetrator. 

For example, users Alice and Eve start chatting on an online dating platform. They move from the dating platform to a social messaging platform, where they exchange explicit pictures~\cite{sextortion_1}. After the encounter, Eve demands that Alice pay them, else they will publish explicit pictures of them. This is based on the findings of Edwards and Hollely~\cite{sextortion_1}, where the majority of sextortion victims report contact with their perpetrator on at least two platforms. 

\medskip
\noindent\textbf{With generative AI}, the only requirement to generate intimate images of a specific person is having a reference photo, meaning that the interaction described in the first phase is no longer required. The only other necessary condition is access to the victim, which can be initiated through public social media accounts. This shift in modus operandi has already been noted by Wei et al.~\cite{wei24ibsahelp}. The interaction between the victim and the perpetrator is replaced by an interaction between the generative AI service and the perpetrator. However, in order to access traces of that interaction such as payment records, the investigation already needs to have identified a potential perpetrator.

In our example, Eve can use publicly available images of Alice on social media to generate the explicit imaginary, or modify already existing material to be explicit. Once the material is generated, they can skip the first phase, and directly extort Alice.

\medskip
\noindent\textbf{Consequences:}
The transformation of the initial phase of sextortion makes it harder to identify the perpetrator in two key ways.

\textit{Amount of interaction:} Duration and the frequency of interactions between perpetrator and survivor are shorter and less frequent. Fewer interactions mean fewer opportunities for the offender to ``slip up'', de-anonymize, or incriminate themselves.

\textit{Location of interaction:} Interactions between survivor and offender no longer span multiple platforms. 
In the described \textit{``before''} example, the survivor and the perpetrator engage with each other on multiple platforms. A decrease in utilized platforms means that there are less places with potential information about the perpetrator. Furthermore, different platforms require and store different information at registration about their users, which can be cross-referenced and regrouped in order to de-anonymize perpetrators. These sources of (partially) identifying information about the perpetrator are no longer available if the only interaction between survivor and perpetrator is extortion. 

To our knowledge, the next stage of the modus operandi where the perpetrator requests payment, is not affected by the advent of generative AI.

\subsection{Voyeurism}\label{sec:voyeurism}
Perpetrators can take photos of a (clothed) survivor without their consent, often focusing on different sexualized areas of the body. This can manifest as up-skirting, down-blousing, and general ``creepshots'' that do not focus on a specific angle. These offenses can be grouped under the term ``voyeurism''.

\medskip
\noindent\textbf{Before} generative methods, offenders would sneak up on a survivor with a camera, then quickly take a photo %
of their body to obtain images. In some cases, they may instead record a person from further away, zooming in on sexualized areas. In partner abuse cases, hidden cameras may be used to capture images, perpetrating abuse without being in the same physical space at the same time. Some perpetrators may distribute this media with others, although it can also be kept for personal use. In all formats, the perpetrator needs to share a physical space with the victim, and is likely to keep copies of the recordings on their phone or camera.

\medskip
\noindent\textbf{With generative AI}, this offense is split into several forms. The most novel form is to obtain any image of the survivor and then use generative AI to ``nudify'' the image, artificially removing clothing from the photo. This can be used to ``enhance'' the voyeurism materials gathered in the traditional manner, removing the clothing that was present when images were obtained. However, these tools can also be used on \textit{any public image of the victim}. This means that an offender can obtain images from social media that were posted in completely innocent contexts, then remove clothing from a survivor without any risk from approaching them in person.

\medskip
\noindent\textbf{Consequences:}
Generating voyeuristic content has become a low-risk offense. There is no longer a requirement to be in close physical proximity at some point, which eliminates the creation of physical traces associated with the presence of a person in a space (f.ex fingerprints on a camera left in a changing room) or CCTV recordings of the person creating the NCEI. %
If investigators obtain a warrant for the suspected perpetrator's devices, then identify they can still find traces of the non-consensual images stored on the phone, however now in addition there might be traces available associated with the creation of the NCSEI -- the deepfake and nudification applications used and their service providers will have traces of perpetrator queries, uploaded and downloaded images, and metadata such as their IP address. %

\subsection{Cyberflashing}\label{sec:cyberflashing}
Cyberflashing is the offense of sharing unsolicited sexual media to others without consent \cite{wei24ibsahelp}. It often targets women, public figures, and dating app users.

\medskip
\noindent\textbf{Before} perpetrators would create intimate images, usually of themselves, to be distributed. These may be send over social media, through online dating apps that permit sharing images\footnote{In particular, Grindr allows users to message each other without any ``matching'' system as a proxy for interest. In combination with the hookup culture present on the platform, this leads to cyberflashing being a normalized behavior that does not incur consequences.}, and to nearby users through AirDrop. Carefully cropped images, fake social media accounts, and anonymized device names can hide the perpetrator's identity, leading to a consequence-free offense in many cases.

\medskip
\noindent\textbf{With generative AI}, and thus the advent of easily available deepfake technologies, the perpetrator has the possibility to create and send synthetic intimate imagery. This is a minor difference, but could theoretically be used to anonymize oneself.

\medskip
\noindent\textbf{Consequences:}
If the perpetrator chooses to send images of themselves or reuses the images, there is a way to link different cases of cyberflashing based on the images. This would constitute the use of the imagery to create intelligence. However, if the perpetrator chooses to always use images of others or uses synthetic images no information that can be gathered about the perpetrator. This is a type of IBSA that is impacted very little by generative technologies, because the traces most likely to identify the perpetrator are linked with the distribution of the material, not the content itself.

\section{Discussion}\label{sec:discussionfuturework}
Based on four different scenarios, we explored the impact of generative AI on the traces created in IBSA offenses, and thus on the type of forensic reasoning they might serve. As illustrated in the scenarios discussing NCEI (see Section~\ref{sec:revengeporn}) and cyberflashing (see Section~\ref{sec:cyberflashing}), synthetic explicit imagery carries less personally identifiable information about the perpetrator than non-generated types of explicit imagery. Furthermore, generative tools can transform the modus operandi of offenses (see Sections~\ref{sec:sextortion} and \ref{sec:voyeurism}) in a way that minimizes the duration and diversity of locations where the victim and perpetrator interact. This in turn also limits the creation of traces carrying partially identifiable information about the perpetrator, which limits the use of traces for identifying potential suspects. In other words, generative AI makes it easier for offenders to remain anonymous.

Previous work has already discussed how generative AI impacts the scale of IBSA. Now, anyone whose visual data is available is a potential victim of NCSEI~\cite{deepfakerealharm}. Synthetic explicit material is quickly and cheaply generated, which makes experts fear that generative AI technology helps attackers outpace the defense mechanisms~\cite{genAITrustSafety}. Furthermore, it has been established that the democratization of deepfake technology as well as the widespread availability of visual data leads to anybody being able to be a potential perpetrator~\cite{deepfakerealharm}. This increases the potential suspects of GAI-IBSA offenses to almost everybody.

However, if a potential suspect is found and investigators gain access to their devices, there are new traces available. The generation of the synthetic explicit imagery, comes with new and different traces stored on the perpetrators devices and on associated servers. As discussed in the scenarios (see Section~\ref{sec:scenarios}), these created traces allow for evaluative reasoning.

As generative AI becomes more realistic and user-friendly, this shift in the suitability of traces for investigation will only further tip the balance in favor of perpetrators.

\bibliographystyle{plain}
\bibliography{usenix2024_SOUPS}

\end{document}